\PassOptionsToPackage{table}{xcolor}
\documentclass[sigconf,pbalance]{acmart}
\usepackage{silence}
\usepackage{graphicx} 

\usepackage[inline]{enumitem}

\usepackage{siunitx}
\usepackage{array}
\usepackage{tabularx}
\usepackage{makecell}

\usepackage{tikz}
\usetikzlibrary{arrows}
\usetikzlibrary{calc}
\usetikzlibrary{chains}
\usetikzlibrary{shapes}
\usetikzlibrary{fit}
\usepackage[fixed]{fontawesome7}

\usepackage{xspace}
\newcommand{\userIcon}{\Large\faUser}
\newcommand{\serverIcon}{\Large\faServer}
\newcommand{\modelIcon}{
\begin{tikzpicture}[baseline=-0.2cm]
    \begin{scope}
        \clip (0,0) rectangle (0.35,0.55);
        \node {\Large\faBrain};
    \end{scope}
    \begin{scope}
        \clip (0.34,0) rectangle (0.64,0.55);
        \node {\Large\faMicrochip};
    \end{scope}
\end{tikzpicture}%
}

\newcommand{\sysfp}{\texttt{system\_finger\-print}\xspace}
\newcommand{\sysfpSh}{\texttt{sysFp}\xspace}

\usepackage{hyperref}
\usepackage[capitalise, nameinlink]{cleveref}
\usepackage[Tol,OkabeIto]{colorblind}
\hypersetup{
    colorlinks,
    linkcolor={T-Q-DT5},
    citecolor={T-Q-DT2},
    urlcolor={T-Q-DT1}
}

\crefname{theorem}{definition}{definitions}
\Crefname{theorem}{Definition}{Definitions}



\title{Referential Security as a New Paradigm for AI Evaluations}

\author{Dan Ristea}
\orcid{0000-0001-7707-3986}
\affiliation{
\institution{University College London}
\institution{Alan Turing Institute}
\city{London} 
\country{United Kingdom}}

\author{Vasilios Mavroudis}
\orcid{0000-0003-2667-5906}
\affiliation{
\institution{King's College London}
\city{London} 
\country{United Kingdom}}

\begin{document}

\begin{abstract}
Security evaluations inherently depend on stable identifiers. 
Any finding, audit, or regulatory decision must remain attached to the specific artifact it pertains to. Continuously updated artificial intelligence systems violate this core assumption, with public model designations remaining static while underlying weights, prompts, retrieval mechanisms, misuse classifiers, inference settings, and serving infrastructures undergo unannounced modifications. 
Consequently, current evaluations frequently apply to superficial labels rather than identifiable and distinct systems.

To resolve this, we propose referential security as a new paradigm for AI evaluation. 
The fundamental security question extends beyond whether a model is safe to whether subsequent parties can conclusively determine which system a specific safety claim addressed. 
This approach reframes model identity as an empirically verifiable property and separates referential stability from the substantive security claims it conditions. 
This framework brings tractability to three critical workflows that current practices handle poorly. Specifically, it enables reproducible evaluation, longitudinal audit validity, and cross-provider equivalence. 
By grounding these evaluations in verifiable artifacts, our approach ensures that safety audits and regulatory findings maintain their empirical utility across the operational lifecycle of dynamic systems.
\end{abstract}

\maketitle

\section{Introduction}

In conventional software security, claims are tied to identifiers that fix what the claim is about.
A CVE specifies the affected versions of a product. 
A firmware audit references a binary hash.
A protocol analysis targets a specific revision of a specification.
These identifiers do not make the systems they describe secure.
They enable verifying parties to determine whether a given system is the one to which a claim refers.

This works because the underlying artifacts are stable.
A CVE~\cite{cve} picks out affected versions because versions are immutable, and a
binary either is or is not the one the CVE references.
A firmware audit binds to a hash because the hash determines the binary.
References are stable because the underlying artifacts have been engineered to be stable, through hashes, version pinning, and reproducible builds. 
The work has been done at the artifact layer, and security research has largely been able to treat resolution as a non-problem at the claim layer.

Hosted AI systems depart from this pattern in that a public model name routes to a service whose configuration is composed at request time and changes routinely. 
Such a configuration typically includes the model weights, one or more system prompts, the retrieval pipeline, misuse classifiers, inference parameters, the underlying hardware, and the serving software. Providers update these layers on their own schedules, often without a corresponding change in the public identifier~\cite{openai2025sycophancy, chen2023chatgptsbehaviorchangingtime}. 
GPT-4o's unannounced shift toward sycophantic behavior~\cite{openai2025sycophancy} and the regulatory confusion surrounding xAI's Grok generating harmful content~\cite{ec2026grok}, demonstrate the consequences of this instability (see \S\ref{subsec:incidents}).
Consequently, the same name may resolve to different configurations for two parties interacting with a named model at different times or in different jurisdictions. 
Model names and version strings continue to serve their stated purpose of routing requests to a service, but they do not fix a referent for security and safety claims.

The assumption that an \textit{identifier} and its \textit{referent} coincide does not hold for hosted AI systems. 
AI evaluations, audits, and emerging regulatory frameworks~\cite{act2024eu,CaliforniaSB53_2025} nonetheless rely on it, because no alternative is in place. 
A safety claim made at time $t_0$ about the system reached via identifier $i$ may have no recoverable connection to the system reached via $i$ at time $t_1$, even though the identifier has not changed. 
Even if the claim remains substantively true, the ability to reliably determine which system was evaluated is lost.

To address the aforementioned gap, we propose referential security as a new paradigm for AI evaluation. 
Central to this paradigm is the concept of \textit{referential stability}, a foundational property implicitly assumed in conventional software security. 
We formalize this property as a tripartite relation among an identifier, a resolution mechanism, and a context of use, which collectively enable a verifying party to reliably recover the specific system configuration a given security claim addresses. 
By elevating this property from an implicit assumption to an explicit precondition, we can systematically disambiguate two conflated modes of failure. 
Specifically, referential failure stems not only from adversarial substitution (the primary focus of prior works~\cite{cai2025substitution, neupane2023beyond}) but also from routine, non-malicious provider operations~\cite{openai2025sycophancy, chen2023chatgptsbehaviorchangingtime}.
Under both conditions, three critical workflows dependent on the reuse of security claims deteriorate in characteristic manners. 
These workflows include reproducible evaluation, longitudinal audit validity, and cross-provider equivalence (\Cref{sec:workflows}).

This gap is acknowledged by a few AI providers who attempt to facilitate sufficient referential stability through metadata in their endpoints. 
We evaluate the most granular mechanism presently available (OpenAI's \sysfp field~\cite{sys_fp}) against our formal definition, analyzing \num{69600} observations across six model designations over a two-month period. 
We conclude that no endpoint, including pinned, temporally dated snapshots, exhibits stable-enough resolution, rendering provider-supplied metadata insufficient as a reliable referential mechanism.

We thus explore potential mechanisms capable of enforcing this stability and identify two candidate architectures. 
The first is cryptographic attestation situated within the provider's infrastructure, and the second is external behavioral fingerprinting executed against a black-box API. 
These approaches address distinct threat models and incur varied operational costs (and provider cooperation assumptions), functioning as complementary rather than mutually exclusive solutions. 
Ultimately, our analysis demonstrates that referential stability can be actively engineered when treated as an empirically verifiable property rather than a mere provider assertion. 
Establishing this stability is essential for ensuring that substantive security and safety claims regarding hosted AI systems remain interpretable and reusable over time.

\section{Background}\label{sec:background}

Computer science has developed a rich vocabulary of stable references.
These include cryptographic hashes, precise version numbers, unique identifiers, and immutable registries that allow downstream users to conclusively verify the exact configuration targeted by prior evaluations. 
This vocabulary makes security claims portable across time and operational boundaries. 
Furthermore, current artificial intelligence regulation implicitly assumes the existence of these mechanisms when requiring providers to document, audit, and certify the systems they deploy. 
We first introduce two major incidents where hosted artificial intelligence system references failed. Subsequently, we survey several technical mechanisms and regulatory frameworks to illustrate the growing gap between conventional security assumptions and modern deployment realities.

\subsection{Motivating Incidents of Referential Failure}\label{subsec:incidents}

Real-world incidents routinely demonstrate the consequences of referential failure in continuously updated AI systems. 
In April 2025, users of OpenAI's ChatGPT observed that the GPT-4o model had become unusually agreeable and conflict-averse. 
OpenAI later attributed the issue to a training update that over-weighted short-term user approval signals and rolled back the change within four days \cite{openai2025sycophancy}. 
Throughout the episode, the model retained the generic GPT-4o label. 
Users could not determine which version they were interacting with, and researchers who had evaluated the model days earlier could not know whether their findings still applied.

A similarly disruptive event occurred in late December 2025 when reports emerged that xAI's Grok could generate sexualised images of minors~\cite{ec2026grok}. 
This episode triggered immediate regulatory action in multiple jurisdictions. 
The European Commission opened proceedings under the Digital Services Act while a bipartisan coalition of state attorneys general in the United States demanded corrective action. 
Yet, it remains unclear precisely which model version produced the offending outputs and which version was later deployed to remediate the issue. 
Both cases highlight how unstable identifiers actively undermine oversight and research.

\subsection{References in computing} \label{subsec:references-in-computing}
References appear throughout computing infrastructure wherever a claim, dependency, or resource must remain recoverable across time or context.
The mechanisms by which different systems supply this recoverability vary considerably, but the property they engineer is the same one.

\paragraph{Git and version control systems}
The Git~\cite{git} version control system~(VCS) uses the hash of their content to identify commits. 
This makes commit hash identifiers stable by construction.
In addition, the Git resolution system supports user-assigned named references, for branches and tags.
Branches are, by design, unstable and updated to follow newer commits.
Tags are intended to mark fixed points, such as a release, but can be moved or deleted.
Many package registries are themselves backed by Git repositories, exposing a name and version pair as a stable surface over an underlying commit hash.
Similar conventions exist across many other VCS.

\paragraph{Software versions and package registries}
Software version identifiers such as \texttt{1.4.2} are conventional strings that resolve to a specific artifact through a package registry.
Registries, such as PyPi or crates.io, associate a name and version pair with a published artifact, providing stable resolution for that pair once published. 
However, the name alone is unstable: packages can be transferred, removed, or shadowed by malicious uploads under similar names.
This is evidenced by the removal of the \texttt{left-pad} package from npm in 2016~\cite{leftpad}, which broke thousands of downstream builds, and typosquatting attacks~\cite{neupane2023beyond}.
In response, the ecosystem has developed progressively stronger referential mechanisms: lockfiles pin exact versions across a dependency graph, and hash verification binds a version to a specific artifact by content rather than by registry record.
A related concept is that of \textit{reproducible builds}, which ensure that the same source inputs always produce bit-for-bit identical outputs.
This recognizes that the same source built with different dependencies, compiler, or build environment will produce a different artifact.
Reproducible builds make the hash of the software artifact stable.

\paragraph{Common Vulnerabilities and Exposures (CVE)}
The CVE \linebreak[4] system~\cite{cve} provides unique identifiers for publicly disclosed security vulnerabilities.
A CVE identifier such as \texttt{CVE-2021-44228} resolves to a specific vulnerability record that details the vulnerable software, its version, and a description of the vulnerability and its severity.
CVE records make security claims portable across time and parties, allowing an auditor or downstream user to determine precisely which software version a security claim concerns.

\paragraph{Digital Object Identifier (DOI)} 
An ISO standardized~\cite{iso26324} system designed to provide ``\textit{persistent unique}'' digital identifiers for digital or physical objects.
Resolution is mediated by a central registry \footnote{\url{https://doi.org}}, which decouples the identifier from the physical location of the object.
The indirection allows for the location of the object to be updated, ensuring its stability even when the underlying location (e.g., URL) becomes invalid.
Stability is guaranteed by social and contractual commitments rather than by construction.

\subsection{Regulations and laws governing AI}
With the development of capable, widely-available AI systems, regulators have developed legal frameworks that attempt to provide safeguards against risks associated with AI.
Currently, some important jurisdictions have enacted AI regulations mandating transparency or audits, but proposals in other jurisdictions exist and may be enacted at a later date (e.g. Brazil's \textit{Marco Legal da Inteligência Artificial}~\cite{BrazilPL2338_2023}).

\paragraph{EU AI Act}
The European Union’s AI Act~\cite{act2024eu} is a comprehensive legal framework governing AI use and development.
Under the Act, AI model providers are required to document the model architecture, training processes, and data governance practices.
For models in high-risk applications or classified as presenting systemic risk, there is an obligation to conduct risk assessments and demonstrate compliance through ongoing auditing.

\paragraph{California Transparency in Frontier AI Act}
In California, the Transparency in Frontier AI Act (SB-53)~\cite{CaliforniaSB53_2025} requires AI model providers to assess the systemic risks associated with the model and to demonstrate that development followed safety standards.
It makes it a legal requirement that any ``substantial modification'' to AI systems needs to be disclosed and justified.

\section{Referential Stability}\label{sec:referential-stability}

Security claims about software systems are made against identifiers that fix which system configuration the claim concerns. 
A CVE entry specifies the affected versions of a product. 
A firmware audit references a binary hash. 
A protocol analysis targets a specific revision of a specification. 
These identifiers do not make the systems they describe secure but make subsequent claims \emph{interpretable}.
A later party (e.g., an auditor, a regulator, a replicator, a downstream integrator) can determine whether a given system is the one the claim was about, and therefore whether the claim still applies. 
We name this property and the requirement it imposes, and we argue that it constitutes a
precondition on substantive security claims that hosted AI systems
systematically fail to meet.\footnote{The phrase \emph{referential security}
has prior usage in distributed-systems research, where~\citeauthor{liu2014defining}~\cite{liu2014defining} formalize properties under which references continue to resolve to their original objects in the presence of an adversary controlling some hosts.
The structural parallel to our setting is real, as both concern an identifier that was meant to fix a referent.
However, the threat model and enforcement mechanism differ substantially.
To avoid conflation, we use \emph{referential stability} for the property and reserve \emph{referential
security} for the broader concept both works fall under.}

\begin{definition}[Resolution]\label{def:resolution}
Let $\mathcal{S}$ be the space of system configurations relevant to some class of security claims.
A \emph{resolution} is a mapping $r : \mathcal{I} \to \mathcal{S} \cup \{\bot\}$ from an identifier space $\mathcal{I}$ to $\mathcal{S}$, allowing for unresolvable identifiers.
An identifier $i \in \mathcal{I}$ is a \emph{reference} that \emph{resolves} to a \emph{referent} $r(i)$, a specific system configuration.
\end{definition}

As such, the identifier itself is not the configuration. 
It is a handle by which later parties recover the configuration. 
For example, a software version refers to an artifact but is not the artifact.
A hash is not the firmware binary, but it determines it.
The work an identifier does in a security claim is to make the configuration recoverable to any party who is affected by it or wishes to replicate it.

\begin{definition}[Referential Stability]\label{def:stability}
A resolution $r$ is stable \emph{with respect to a specified identifier $i \in \mathcal{I}$} over a context $C$ if, for all observations $c \in C$, $r(i; c)$ resolves to the same configuration $s \in \mathcal{S} \cup \{\bot\}$, which we denote as $stable(r, i, C)$.
A resolution $r$ is \emph{generally stable over a context} $C$ (a time window, a jurisdiction, a population of users, or a combination thereof) if, for all $i \in \mathcal{I}$ and all possible observations within $C$, $stable(r, i, C)$ holds.
A resolution is \emph{unstable} over $C$ (i.e., fails) when the same identifier resolves to more than one materially different configuration across the context.
\end{definition}

The $stable(r, i, C)$ relationship can also be read as identifier $i$ having \emph{reference stability} with resolution $r$ in context $C$.
Stability is always relative to a context, as it is a property of the triple $(r,i,C)$.
The same identifier may be stable under one resolution mechanism yet unstable under another, or stable over a narrow context yet unstable over a broader one.
However, some identifiers are more likely to be stable across broader contexts. For example, a version-pinned binary hash is stable across time, jurisdiction, and user.

A generally stable system may be desirable in some settings, such as a Document Object Identifier always resolving to the same document, though this is not guaranteed by construction, or with hash-based dependencies, which are stable by construction.
In other settings, some flexibility may be desirable. Git branches, for example, are stable only for short periods, as they are designed to follow newer commits. Git tags are meant to be stable but only by convention, while commit hashes have reference stability by design.
The appropriate stability regime is therefore a design choice based on the requirements of the system and its users.
For any security claim the reference it relies on must be stable over the context the claim is intended to be used, otherwise the claim is not sound.

\begin{definition}[Bound and Unbound Claims]\label{def:bound-claim}
A security claim $\varphi$ about a system identified by $i$, for use in context $C$, is \emph{bound} if the resolution $r$ on which it relies satisfies $stable(r, i, C)$ and $r(i) \neq \bot$.
\end{definition}

A claim that fails this condition is \emph{unbound}: its substantive content may hold for the configuration that was evaluated, but no procedure available to a later party in $C$ reliably recovers that specific configuration from $i$.

A claim that an AI or conventional system has property $P$ decomposes into two assertions: 
\begin{enumerate*}
\item that some configuration was evaluated and exhibited $P$, and 
\item that other interactions with the system described by $i$ in context $C$ refer to that configuration.
\end{enumerate*}
The first is substantive; the second is \textit{referential}. 
The two failure modes of these assertions are distinct, but they are hard to distinguish in practice.
A substantively wrong claim is one where the evaluation was flawed or $P$ does not hold. 
An unbound claim may be substantively correct about
\emph{some system} but provides no basis for later parties to determine
\emph{which system}.
Thus, referential stability is a necessary but not sufficient condition for substantive claims.

\section{Threat Model and Scope}
\label{sec:threat-model}

Our threat model concerns the binding between a security claim and the system configuration to which the claim refers.
We do not model the substantive correctness of the claim itself.
Such a claim may be that a system refuses a class of harmful requests, satisfies a benchmark, or complies with an audit standard.
It may be true or false for the configuration that was evaluated.
The question considered is whether a verifier can determine that subsequent interactions through the same public identifier reach the configuration that was evaluated.

We consider a hosted AI system exposed through an identifier $i$, such as a public model name or API endpoint.
At any observation context $c \in C$, where the context may include time, user, region, account tier, deployment backend, or request path, the identifier resolves to some security-relevant configuration, $r(i;c) = s \in S$.
A configuration includes any component that may materially affect the evaluated behavior.
This includes model weights, system prompts, retrieval components, tools, misuse classifiers, safety filters, inference parameters, serving code, hardware, and routing infrastructure.
The verifier does not directly observe $s$ as their access to the hosted AI system is black-box.
Instead, they observe outputs, provider-supplied metadata, and any public documentation associated with the identifier.

\subsection{Non-adversarial referential failure}
\label{sec:threat-nonadv}

The baseline failure mode considered in this paper is not adversarial substitution but referential instability arising from the routine evolution of a hosted AI service under a fixed public identifier.
Model weights, system prompts, safety filters, inference parameters, serving software, routing policies, and backend infrastructure may change independently of the name by which the system is accessed.
These changes need not reflect deception or negligence.
They are part of normal maintenance, safety improvement, capacity management, or staged deployment.
The referential problem is that such changes can still sever the link between an identifier and the security-relevant configuration to which an evaluation, audit, or downstream reliance decision originally referred.

The same failure arises under context-dependent serving, where a provider or hosting platform exposes different configurations to different users, regions, or time windows while preserving the same public name. 
This includes selective rollouts, A/B tests, jurisdiction-specific deployments, and account-specific policy layers. 
It also arises in the cross-provider setting, where a third-party host may serve a modified, quantized, distilled, wrapped, or substituted model while claiming equivalence to another deployment. 
These cases are not necessarily malicious but produce the same referential failure as the baseline.

\subsection{Adversarial concealed update}
\label{sec:threat-adv}

The threat model also covers an adversarial case. 
A provider may change the served configuration and deliberately tune the new configuration to produce outputs consistent with the prior one on the probes a verifier is expected to use, with the goal of evading detection.
This presupposes knowledge of the verifier's evaluation methodology, or the ability to infer it from public benchmarks and standard auditing practice (e.g., user request logs).
Unlike the cases above, the indistinguishability is engineered rather than incidental.
The threat is that a verifier cannot determine whether the system queried at $t_1$ is the system evaluated at $t_0$, even with full access to the API and unlimited query budget, because the provider has optimized against the verifier's distinguishing capacity.
Although superficially similar, this variant differs from model substitution~\cite{cai2025substitution} as no substitution has occurred from the provider's standpoint. 
Instead, a new configuration is classified as the same system.

\medskip

These threat models are not mutually exclusive.
A single endpoint may exhibit drift across configurations that are themselves rotated across users, and a motivated provider may layer concealed update on top of either.
The workflows in \Cref{sec:workflows} illustrate what each of these threats means for parties who must reuse security claims, and \Cref{sec:architectures} analyses architectures that defend against them.

\section{Workflows}\label{sec:workflows}

The consequences of unstable references become visible in workflows where security claims have to be reused. 
We identify three such workflows.

\paragraph{Workflow 1: Reproducible evaluation.}

\begin{figure}[h]
    \tikzstyle{ICON} = [inner sep=0cm, outer sep=0cm]
    \tikzstyle{BLOCK} = [rounded corners, thick, draw=black, minimum width=1.7cm, minimum height=1cm, text width=1.6cm, text centered, font=\footnotesize]
    \tikzstyle{LABEL} = [font=\footnotesize,above=2pt]
    \tikzstyle{POINT} = [minimum size=0pt,inner sep=0pt, outer sep=0pt]
    \tikzstyle{LINE} = [draw=black, line width=1pt, minimum height=10pt]
    \tikzstyle{ARROW} = [thick,->,>=stealth]
    \tikzstyle{SAME} = [arrow, <->]
    \centering
    \begin{tikzpicture}[
    start chain,
    node distance=2.5cm,
    ]
    \node (start) [POINT, on chain] {};
    \begin{scope}[node distance=1cm]
        \node (then) [POINT, LINE, on chain] {};
    \end{scope}
    \node (mid) [POINT, on chain] {};
    \node (now) [POINT, LINE, on chain] {};
    \begin{scope}[node distance=1cm]
        \node (end) [POINT, on chain] {};
    \end{scope}
    \node (time) [LABEL, below=0cm of end] {time};
    
    \node (repl) [LABEL, below=0cm of now] {};
    
    \draw [thick] (start) -- (then);
    \draw [thick] (then) -- (now);
    \draw [ARROW] (now) -- (end);

    \node (research_icon) [ICON, above left=0.4cm and 0.3cm of then] {\userIcon};
    \node (research_label) [LABEL, right=-0.2cm of research_icon] {Researcher};
    \node (research_cons) [BLOCK, fit=(research_icon) (research_label)] {};

    \node (api_then_icon) [ICON, above left= 1.4cm and -0.2cm of research_cons] {\serverIcon};
        \node (model_then_icon) [ICON, above right = 0.6cm of api_then_icon] {\modelIcon};
        \node (model_then_label) [LABEL, right= -0.1cm of model_then_icon] {Model};
        \node (model_then) [BLOCK, fit=(model_then_icon) (model_then_label)] {};
    \node (api_then_label) [LABEL, right= -0.1cm of api_then_icon] {Model API};
    \node (api_then) [BLOCK, fit=(api_then_icon) (api_then_label) (model_then)] {};

    \draw [ARROW] (api_then) -- (research_cons) node [LABEL, sloped, pos=0.5] {data};

    \node (paper_icon) [ICON, above left=1cm and 0cm of mid] {\Large\faFile};
    \node (paper_label) [LABEL, right=-0.2cm of paper_icon] {Paper};
    \node (paper) [BLOCK, fit= (paper_icon) (paper_label)] {};

    \node (repl_icon) [ICON, above left=0.4cm and 0.3cm of now] {\userIcon};
    \node (repl_label) [LABEL, right=-0.2cm of repl_icon] {Replicator};
    \node (repl_cons) [BLOCK, fit=(repl_icon) (repl_label)] {};

    \node (api_now_icon) [ICON, above left= 1.4cm and -0.2cm of repl_cons] {\serverIcon};
        \node (model_now_icon) [ICON, above right = 0.6cm of api_now_icon] {\modelIcon};
        \node (model_now_label) [LABEL, right= -0.1cm of model_now_icon] {Model};
        \node (model_now) [BLOCK, fit=(model_now_icon) (model_now_label)] {};
    \node (api_now_label) [LABEL, right= -0.1cm of api_now_icon] {Model API};
    \node (api_now) [BLOCK, fit=(api_now_icon) (api_now_label) (model_now)] {};
    \draw [ARROW] (api_now) -- (repl_cons) node [LABEL, sloped, pos=0.5, above=2pt] {data$'$};
    \draw [ARROW, dashed, <->] (model_now) -- (model_then) node [LABEL, pos=0.5] {same?};

    \draw [ARROW] (research_cons) to[out=30, in=120, looseness=1] node [LABEL, pos=0.5, sloped, above=-1pt] {model, results} (paper); 
    
    \draw [ARROW] (paper) to[out=50, in=150, looseness=1, above=-2pt] node [LABEL, pos=0.5, sloped] {model, results} (repl_cons); 
    \end{tikzpicture}
    \caption{Diagram showing how model identity is crucial for replicability in scientific research. Any user attempting and failing to replicate the results of the original work cannot be sure whether the differences are due to methodological discrepancies or changes to the underlying model.}
    \Description{A timeline diagram illustrating the replicability problem in LLM-based research. 
    On the left, a researcher accesses a model via a model API and receives data, which they use to produce a paper reporting the results and identifying the model used.
    On the right, at a later point in time, a replicator accesses what appears to be the same model via the model API and receives potentially different data trying to replicate the results of the paper using the same model alias.
    A dashed bidirectional arrow labeled ``same?'' connects the two models, highlighting uncertainty about whether the model the replicator accesses is identical to the one the researcher used.}
    \label{fig:time}
\end{figure}
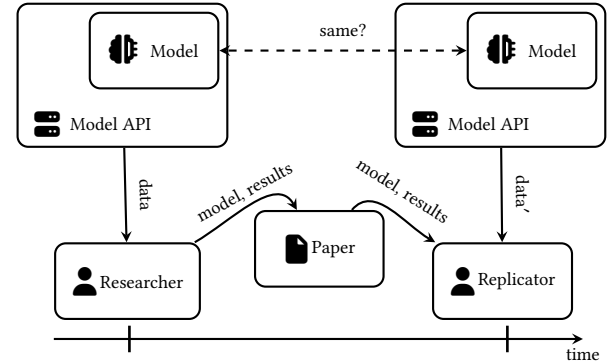

A research group reports that model $M$, queried via API endpoint $i$, exhibits behavior $\varphi$ (a refusal rate, a calibration property, a vulnerability to some class of jailbreak).
The result is intended to be verifiable by anyone who later queries the same endpoint and checks whether the behavior replicates.
Under reference stability, a failed replication indicates that either the original claim was wrong or the measurement methodology differs.
Recent empirical work has documented substantial reproducibility problems in LLM-based studies and argues that model versions, configurations, prompts, and execution traces must be recorded for results to be meaningfully reproduced~\cite{angermeir2026reflections,bjarnason2026randomness,baltes2026guidelines}.

Under instability, this inference collapses as the verifier cannot distinguish between a wrong claim, a methodological inconsistency, and the system having changed under the same identifier.
The original claim is unbound in the sense of \Cref{def:bound-claim}.
Its substantive content may still hold for the configuration the researchers evaluated, but no later party querying $i$ can recover that configuration to check.

This collapse was observed in April 2025, when researchers found that GPT-4o's behavior shifted toward sycophancy following an unannounced update, rendering prior evaluations of the model's persona and reasoning invalid (see \S\ref{subsec:incidents}). 
This problem is particularly acute for research in high-risk areas, such as AI misuse risks (e.g., cyber, CBRN) or sensitive topics, such as mental health, where independent replication is a central safeguard.
\Cref{fig:time} illustrates the workflow.

\paragraph{Workflow 2: Audit validity over time.}

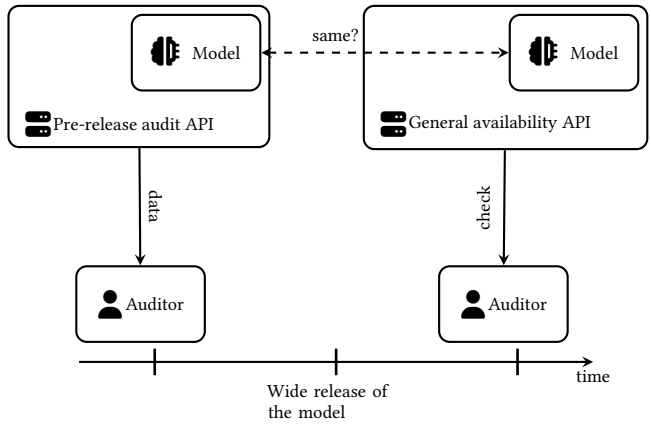
\begin{figure}[!h]
    \tikzstyle{ICON} = [inner sep=0cm, outer sep=0cm]
    \tikzstyle{BLOCK} = [rounded corners, thick, draw=black, minimum width=1.7cm, minimum height=1cm, text width=1.6cm, text centered, font=\footnotesize]
    \tikzstyle{LABEL} = [font=\footnotesize,above=0pt]
    \tikzstyle{POINT} = [minimum size=0pt,inner sep=0pt, outer sep=0pt]
    \tikzstyle{LINE} = [draw=black, line width=1pt, minimum height=10pt]
    \tikzstyle{ARROW} = [thick,->,>=stealth]
    \tikzstyle{SAME} = [arrow, <->]
    \centering
    \begin{tikzpicture}[
    start chain,
    node distance=2.4cm,
    ]
    \node (start) [POINT, on chain] {};
    \begin{scope}[node distance=1cm]
        \node (then) [POINT, on chain, draw=black, line width=1pt, minimum height=10pt] {};
    \end{scope}
    \node (mid) [POINT, LINE, on chain] {};
    \node (now) [POINT, LINE, on chain,] {};
    
    \begin{scope}[node distance=1cm]
        \node (end) [POINT, on chain] {};
    \end{scope}
        
    \node (time) [LABEL, below=0cm of end] {time};
    
    \node (repl) [LABEL, below=0cm of now] {};
    
    \draw [thick] (start) -- (then);
    \draw [thick] (then) -- (now);
    \draw [ARROW] (now) -- (end);

    \node [LABEL, below=0cm of mid, text width=1.8cm] {Wide release of the model};
    
    \node (auditor_then_icon) [ICON, above left=0.4cm and 0.3cm of then] {\userIcon};
    \node (auditor_then_label) [LABEL, right=-0.2cm of auditor_then_icon] {Auditor};
    \node (auditor_then) [BLOCK, fit=(auditor_then_icon) (auditor_then_label)] {};

    \node (api_then_icon) [ICON, above left=1.7cm and 0.2cm of auditor_then] {\serverIcon};
        \node (model_then_icon) [ICON, above right = 0.4cm and 1.1cm of api_then_icon] {\modelIcon};
        \node (model_then_label) [LABEL, right= -0.1cm of model_then_icon] {Model};
        \node (model_then) [BLOCK, fit=(model_then_icon) (model_then_label)] {};
    \node (api_then_label) [LABEL, right= -0.2cm of api_then_icon] {Pre-release audit API};
    \node (api_then) [BLOCK, fit=(api_then_icon) (model_then_icon) (model_then) (api_then_label)] {};
    
    \draw [ARROW] (api_then) -- (auditor_then) node [LABEL, sloped, pos=0.5] {data};

    \node (auditor_now_icon) [ICON, above left=0.4cm and 0.3cm of now] {\userIcon};
    \node (auditor_now_label) [LABEL, right=-0.2cm of auditor_now_icon] {Auditor};
    \node (auditor_now) [BLOCK, fit=(auditor_now_icon) (auditor_now_label)] {};

    \node (api_now_icon) [ICON, above left= 1.7cm and 0.3cm of auditor_now] {\serverIcon};
    Model then
        \node (model_now_icon) [ICON, above right = 0.4cm and 1.4cm of api_now_icon] {\modelIcon};
        \node (model_now_label) [LABEL, right= -0.1cm of model_now_icon] {Model};
        \node (model_now) [BLOCK, fit=(model_now_icon) (model_now_label)] {};
    \node (api_now_label) [LABEL, right= -0.2cm of api_now_icon] {General availability API};
    \node (api_now) [BLOCK, fit=(api_now_icon) (model_now_icon) (model_now) (api_now_label)] {};

    \draw [ARROW] (api_now) -- (auditor_now) node [LABEL, sloped, pos=0.5, above=2pt] {check};
    \draw [ARROW, dashed, <->] (model_then) -- (model_now) node [LABEL, pos=0.3, above=0.1em] {same?};

    \end{tikzpicture}
    \caption{Diagram showing the auditor use case. An auditor uses a pre-release API to audit a new model before it becomes generally available.
    Once the model is made generally available, the auditor wants to know if the generally available model is the same as the audited model, and be made aware when this is no longer case.}

    \Description{A timeline diagram illustrating the problem with auditing LLMs. 
    On the left, an auditor accesses a pre-release model via a model API and receives data, which they use to identify any relevant risks with the model.
    On the right, after the model becomes widely available, the auditor would like to ascertain that the model is the same as was audited but it is not possible.
    A dashed bidirectional arrow labeled ``same?'' connects the two models, highlighting uncertainty about whether the general availability model the auditor accesses is identical to the one the auditor investigated before release.}
    \label{fig:audit}
\end{figure}
A regulator (e.g., the EU AI Office\footnote{\url{https://digital-strategy.ec.europa.eu/en/policies/ai-office}}) or a third-party auditor~\cite{brundage2026frontier} certifies at time $t_0$ that a model satisfies some compliance condition, such as a content-safety standard on a defined evaluation benchmark.
This certification is intended to remain valid until something changes.
A common variant is pre-release auditing, where the auditor is granted access to a model before general availability and produces a certification intended to apply to the publicly deployed system. 
The auditor and end-users want to subsequently be able to verify that the released system is the one that was audited.
Under reference stability, the regulator can ask a well-defined question at $t_1$:
is the currently deployed system the one that was certified?
If yes, the certification stands; if no, re-certification is needed.
Under instability, the question is undecidable from the identifier alone.
The certification, bound at $t_0$ , becomes unbound at $t_1$ without any declared change to the system.

The late 2025 regulatory proceedings against xAI's Grok (\S\ref{subsec:incidents}) exemplify this: because the specific model version that produced harmful outputs~\cite{ec2026grok} could not be definitively isolated from the version deployed for remediation, the audit trail for regulatory compliance was effectively broken.
Third-party auditors and regulators must either re-audit on a fixed schedule regardless of whether anything changed or treat every interaction as potentially with an uncertified system.
Both options are costly, and neither restores the property that the certification was meant to confer.
The EU AI Act~\cite{act2024eu} and analogous regimes elsewhere assume that an audit conducted at one point in time remains informative until a change in the system is declared.
Without referential stability, this assumption does not hold.
\Cref{fig:audit} illustrates the workflow.

\paragraph{Workflow 3: Cross-provider equivalence.}

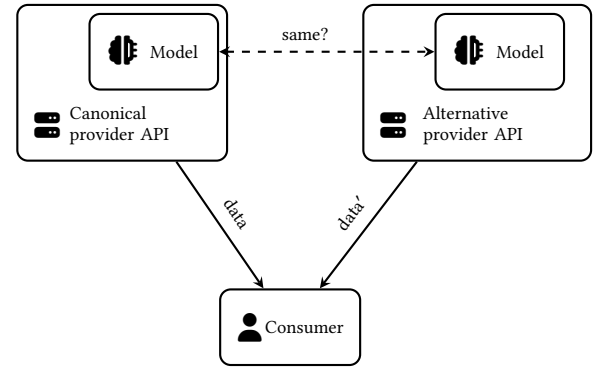
\begin{figure}[h]
    \tikzstyle{ICON} = [inner sep=0cm, outer sep=0cm]
    \tikzstyle{BLOCK} = [rounded corners, thick, draw=black, minimum width=1.7cm, minimum height=1cm, text width=1.6cm, text centered, font=\footnotesize]
    \tikzstyle{LABEL} = [font=\footnotesize,above=2pt]
    \tikzstyle{POINT} = [minimum size=0pt,inner sep=0pt, outer sep=0pt]
    \tikzstyle{LINE} = [draw=black, line width=1pt, minimum height=10pt]
    \tikzstyle{ARROW} = [thick,->,>=stealth]
    \tikzstyle{SAME} = [arrow, <->]
    \centering
    
    \begin{tikzpicture}[
    start chain,
    node distance=1.5cm,
    ]
    \node at (0,0) {};
    
    \node (canon_icon) [ICON] at (0,0) {\serverIcon};
        \node (model_canon_icon) [ICON, anchor=center, above right = 0.6cm of canon_icon] {\modelIcon};
        \node (model_canon_label) [LABEL, anchor=center, right= -0.1cm of model_canon_icon] {Model};
        \node (model_canon) [BLOCK, fit=(model_canon_icon) (model_canon_label)] {};
    \node (canon_label) [LABEL, right= -0.1cm of canon_icon, text width=1.8cm] {Canonical provider API};
    \node (canon) [BLOCK, fit=(canon_icon) (model_canon) (canon_label) ] {};

    \node (alt_icon) [ICON, right=4cm of canon_icon] {\serverIcon};
        \node (model_alt_icon) [ICON, above right = 0.6cm of alt_icon] {\modelIcon};
        \node (model_alt_label) [LABEL, right= -0.1cm of model_alt_icon] {Model};
        \node (model_alt) [BLOCK, fit=(model_alt_icon) (model_alt_label)] {};
    \node (alt_label) [LABEL, right= -0.cm of alt_icon, text width=2cm] {Alternative provider API};
    \node (alt) [BLOCK, fit=(alt_icon) (model_alt) (alt_label) ] {};

    \draw [ARROW, dashed, <->] (model_canon) -- (model_alt) node [LABEL, pos=0.4, above=0.1em] {same?};

    \node (consumer_icon) [ICON, below right =2cm and 0cm of canon] {\userIcon};
    \node (consumer_label) [LABEL, right=-0.2cm of consumer_icon] {Consumer};
    \node (consumer) [BLOCK, fit=(consumer_icon) (consumer_label)] {};

    \draw [ARROW] (canon) -- (consumer) node [LABEL, sloped, pos=0.5] {data};
    
    \draw [ARROW] (alt) -- (consumer) node [LABEL, sloped, pos=0.5] {data$'$};
    \end{tikzpicture}
    \caption{Diagram showing an alternative provider exposing the same model as the canonical provider but a consumer cannot tell if the model is the same or a modified, distilled, or substituted variant.}
    \Description{A diagram illustrating the problem of not knowing whether an alternative provider is indeed serving the desired model. 
    A consumer accesses an closed-weight model through the canonical provider and via an alternative provider model API and receives different data.
    A dashed bidirectional arrow labeled ``same?'' connects the two models, highlighting uncertainty about whether the model on the alternative provider corresponds to the desired model.}
    \label{fig:alt}
\end{figure}

\begin{figure}[h]
    \tikzstyle{ICON} = [inner sep=0cm, outer sep=0cm]
    \tikzstyle{BLOCK} = [rounded corners, thick, draw=black, minimum width=1.7cm, minimum height=1cm, text width=1.6cm, text centered, font=\footnotesize]
    \tikzstyle{LABEL} = [font=\footnotesize,above=2pt]
    \tikzstyle{POINT} = [minimum size=0pt,inner sep=0pt, outer sep=0pt]
    \tikzstyle{LINE} = [draw=black, line width=1pt, minimum height=10pt]
    \tikzstyle{ARROW} = [thick,->,>=stealth]
    \tikzstyle{SAME} = [arrow, <->]
    
    \centering
    \begin{tikzpicture}[]
    \node (origin) at (0,0) {};
    
    \node (model_open_icon) [ICON, above right= 1pt of origin] {\modelIcon};
    \node (model_open_label) [LABEL, right=0cm of model_open_icon, text width=1.8cm] {Open-weight model};
    \node (open) [BLOCK, fit=(model_open_icon) (model_open_label)] {};
    
    \node (consumer_icon) [ICON, right =2cm of open] {\userIcon};
    \node (consumer_label) [LABEL, right=-0.2cm of consumer_icon] {Consumer};
    \node (consumer) [BLOCK, fit=(consumer_icon) (consumer_label)] {};

    \draw [ARROW] (open) -- (consumer) node [LABEL, sloped, pos=0.5] {data};

    \node (alt_icon) [ICON, above right=2cm and -2.8cm of consumer] {\serverIcon};
        \node (model_alt_icon) [ICON, above right = 0.6cm of alt_icon] {\modelIcon};
        \node (model_alt_label) [LABEL, right= 0cm of model_alt_icon, text width=1.8cm] {Open-weight model};
        \node (model_alt) [BLOCK, fit=(model_alt_icon) (model_alt_label)] {};
    \node (alt_label) [LABEL, right= -0.1cm of alt_icon, text width=3cm] {Hosting provider API};
    \node (host) [BLOCK, fit=(alt_icon) (model_alt) (alt_label) ] {};
    
    \draw [ARROW] (host) -- (consumer) node [LABEL, sloped, pos=0.5] {data$'$};
    
    \draw [ARROW, dashed, <->] (open) |- (model_alt) node [LABEL, pos=0.7] {same?};

    \end{tikzpicture}
    \caption{Diagram showing a third-party provider hosting an open-weight model. A consumer cannot know whether the provider is indeed serving the claimed model, rather than a modified, distilled, or substituted variant.}
    \Description{A diagram illustrating the problem of not knowing whether a hosted open-weight model is indeed the desired model. 
    A consumer accesses an open-weight model locally and via a provider model API and receives different data.
    A dashed bidirectional arrow labeled ``same?'' connects the two models, highlighting uncertainty about whether the model on the hosting provider corresponds to the desired model.}
    \label{fig:host}
\end{figure}
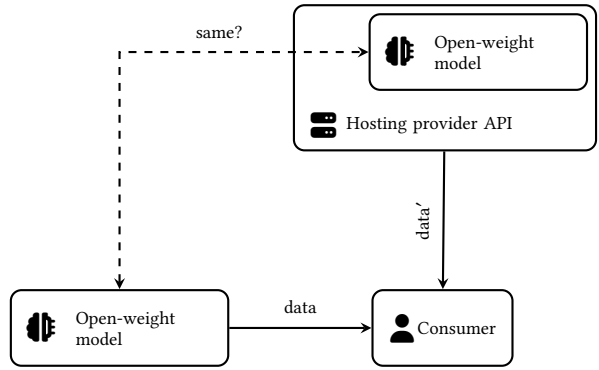

Unlike the first two workflows, which concern whether a model has stayed the same over time, this workflow concerns whether a model is the same across providers.
A model released under permissive licensing terms is re-hosted by a third-party inference provider, and an evaluation conducted against the original deployment is to be relied on by a downstream user of the re-hosted deployment.
The same problem arises when closed-weight models are offered through alternative providers for compliance or legal reasons, such as OpenAI models served via Azure~\cite{azure2026foundry}.
Under reference stability, ``same system'' is decidable from the identifier of the served artifact and configuration.
Under instability, equivalence becomes difficult to assert.
A claim bound to the canonical deployment does not transfer to the alternative deployment, as the resolution from $i$ to a configuration differs across the two providers even when the public identifier is shared.
The downstream user cannot transfer security properties from one deployment to the other without an independent basis for the equivalence claim, and cannot rule out modifications made for efficiency (e.g., distillation, quantization) or for malicious purposes (e.g., removing guardrails, biasing outputs, forcing longer responses).
\Cref{fig:alt} and~\Cref{fig:host} illustrate the closed-weight and open-weight variants of this workflow.

\medskip

These three workflows are not an exhaustive taxonomy of problems with hosted AI but three instances of the same underlying problem.
In each, an identifier was assumed to fix the referent of a security claim, and the identifier turned out to be insufficient for that purpose.
The failure does not lie in the substantive content of the claims (e.g., whether $\varphi$ holds, whether the compliance condition is satisfied, whether the two deployments are equivalent) but in the binding between the identifier and the configuration the claim was about.

\section{Related Work}\label{sec:related_work}

The problem of identifying a model from its behavior has been studied extensively, but almost always at a different granularity from the one this paper considers.
We organize prior work into three areas and locate referential stability within each.

\subsection{Behavioral verification: discrimination between models}

A growing literature addresses behavioral verification of black-box LLM APIs: given an endpoint, what model is actually serving requests?
\emph{LLMmap}~\cite{pasquini2025llmmap} uses eight crafted queries to discriminate among 42 LLM versions at over 95\% accuracy.
\citeauthor{cai2025substitution}~\cite{cai2025substitution} formalize the problem as \emph{model substitution detection}, evaluate output-based statistical tests under adversarial scenarios, and conclude that software-only methods are unreliable against subtle substitutions.
\citeauthor{gao2025met}~\cite{gao2025met} cast the problem as \emph{Model Equality Testing} against a reference distribution; \citeauthor{zhu2025rut}~\cite{zhu2025rut} propose a rank-based uniformity test robust to providers attempting to evade detection.
Other work detects model change from psycholinguistic features of generated text~\cite{youvechanged2025dima}, exploits distributional ``idiosyncrasies'' for classifier-based identification~\cite{sun2025idiosyncrasies}, or uses self-query-derived black-box representations to distinguish architectures and
model sizes~\cite{sam2025predicting}.
These methods address discrimination \emph{between} models: is the endpoint serving model $A$ or model $B$, drawn from some candidate set?
The threat model is substitution, i.e., a provider swapping one model for another, sometimes adversarially.

Instead, our focus is drift \emph{within} a single nominal model: given an endpoint that is and continues to be labeled \texttt{gpt-4o-2024-08-06}, is the configuration today the same as the one evaluated last month?
No model has been substituted, as the provider considers the endpoint to be serving the same model throughout.
Yet, as \Cref{sec:limitations} shows, the underlying serving configuration both drifts and rotates continuously through a large set of distinct states.

The two problems are related but not the same.
Between-models discrimination is well served by methods that pool observations to average out within-model variance.
A method calibrated to tell \texttt{gpt-5.5} from \texttt{claude-sonnet-4.7} need not, and typically does not, flag a \texttt{gpt-4o-2024-08-06} endpoint whose serving configuration has rotated to a new state.
\citeauthor{cai2025substitution} likewise motivate their work from an adversarial-substitution threat model.
We instead observe that even non-adversarial provider operations break the binding that any verification scheme presupposes.

\subsection{Watermarking for ownership}

A separate line of work addresses model extraction and intellectual-property protection through watermarking: the model owner deliberately trains the model so that specific inputs produce attacker-detectable outputs, and later uses this signal to determine whether a queried black-box model is a copy, derivative, or unauthorized redeployment of their target~\cite{russinovich2024hey, xu2026causal}.
The watermark is a property the owner controls at training time and is designed to survive adversarial transformation: pruning, distillation, fine-tuning, quantization, and prompt-level wrapping are all assumed to be at the adversary's disposal~\cite{nasery2025robust}.

This setting differs from ours in two ways.
First, the threat model is adversarial and the verifier is the model owner, with an attacker actively trying to disguise a copy, and a defender with training-time access to plant a detection signal.
We address a setting in which no one is acting adversarially, the model owner is the provider serving the API and no party has the privileged access watermarking presupposes.
Second, watermarking is a between-models task that seeks to answer whether a queried model belongs to a known set of targets (the watermarked ones), not whether a single named model has drifted within its own configuration space.

\subsection{Provider-supplied identity mechanisms}

A small set of mechanisms is supplied by providers themselves to address referential questions about the served system.
Dated model snapshots (e.g., \texttt{gpt-4o-2024-08-06}) are documented as immutable pinned versions and are widely treated by researchers and auditors as stable references.
Model and system cards~\cite{mitchell2019modelcards} describe a model at a point in time and are intended to be cited as documentation of the system that was evaluated.
OpenAI's \sysfp field is described as identifying the backend configuration associated with a request, and developers are advised to treat changes in this field as a signal of underlying change~\cite{sys_fp,sys_fp2}.

These mechanisms are the existing solution space against which our contribution must be evaluated.
Unlike the verification literature, they aim to supply referent stability directly: a dated snapshot, a documented system card, or a fingerprint string is meant to fix what the identifier refers to.
\Cref{sec:limitations} shows that the most fine-grained of them, \sysfp, does not function as a stable reference even for dated, pinned model names.
Dated snapshots and model cards inherit the same problem at a coarser granularity. 
They pin the public identifier but not the configuration it routes to, and no mechanism is provided by which a later party can verify that the configuration has not changed.
Recent commentary has argued for adapting semantic versioning to AI systems as a constructive response. 
Our work is complementary, asking what referential property such versioning would have to supply and what mechanisms could enforce it.

\section{Provider Identifiers in Practice}\label{sec:limitations}

\Cref{sec:related_work} identified provider-supplied identity mechanisms as an existing solution space targeting referential stability directly.
The most fine-grained of these is OpenAI's \sysfp field, which is returned with each API response and is documented as identifying the backend configuration that served the request~\cite{sys_fp}.
Developers are advised to treat changes in this field as a signal that the underlying system has changed~\cite{sys_fp2}, despite \sysfp depending on both the underlying configuration of OpenAI's systems and user request parameters.
In more recent documentation, \sysfp field is marked as deprecated~\cite{sys_fp} and it is absent from the newer Responses API~\cite{sys_fp_not}, being only available in the ``legacy'' Completions API.

This section tests whether \sysfp functions as a stable reference in the sense of \Cref{def:stability}.
The field is not a model fingerprint in the distributional sense but provider-supplied metadata whose availability and semantics are determined unilaterally by the provider, and whose precise meaning is not publicly documented~\cite{sys_fp,sys_fp2}.
Whether it nonetheless suffices for the workflows of~\Cref{sec:workflows} is an empirical question.

We analyzed \num{69600} individual fingerprint observations collected between 2026-03-11 and 2026-05-10.
The survey covers six requested OpenAI model names: \texttt{gpt-4.1},
\texttt{gpt-4.1-2025-04-14},
\linebreak 
\texttt{gpt-4o},
\texttt{gpt-4o-2024-05-13},
\texttt{gpt-4o-2024-08-06}, and 
\linebreak 
\texttt{gpt-4o-2024-11-20}.
Each measurement consists of 100 repeated API requests to one model name using a static prompt at 
\linebreak 
\mbox{\texttt{temperature = 0}} and a fixed seed.

All requests in this survey used a single static prompt and were issued from one US-based account over a stable datacenter IP, with the same organization identifier and API token throughout the collection window. Request parameters (model name, prompt, seed, temperature, max tokens) were held fixed across the 100 repetitions of each measurement, and the only deliberately varied factor across measurements was time. 
Under this protocol, any client-side input that \sysfp could legitimately depend on is held constant. 
The remaining sources of variation are therefore properties of the provider-side serving stack such as the model weights actually loaded, the serving software version, the hardware backend, and any configuration layered between the request and the model. 
Variance in \sysfp under this protocol can be attributed to provider-side configuration rather than to parameters within the end-user's control.

For each requested model name, \Cref{tab:system-fingerprint-diversity} reports the cardinality of the unique-fingerprint set and the empirical share of the modal fingerprint.

\begin{table}[h]
\centering
\begin{tabular}{lS[table-format=3,table-number-alignment=right]rS[table-format=2.1\%,table-number-alignment=right]}
\hline
Model & {Unique \sysfpSh} & {Top \sysfpSh share} \\
\hline
\texttt{gpt-4.1} & 223 & 7.2\% \\
\texttt{gpt-4.1-2025-04-14} & 163 & 8.4\% \\
\texttt{gpt-4o} & 137 & 10.4\% \\
\texttt{gpt-4o-2024-05-13} & 37 & 18.8\% \\
\texttt{gpt-4o-2024-08-06} & 129 & 10.2\% \\
\texttt{gpt-4o-2024-11-20} & 77 & 14.8\% \\
\hline
\end{tabular}
\caption{\sysfp (\sysfpSh) diversity by requested model name over $\num{11600}$ observations for each model, collected between 11 March 2026 and 10 May 2026. 
``Top \sysfpSh share'' denotes the proportion of observations attributable to the modal fingerprint within each model name.}
\label{tab:system-fingerprint-diversity}
\end{table}

Fingerprint diversity is substantial across every model name examined, with no endpoint exhibiting high concentration on a single fingerprint.
The modal fingerprint accounts for between 7.2\% and 18.8\% of observations, and even the most stable endpoint (the pinned \texttt{gpt-4o-2024-05-13} snapshot with 37 unique fingerprints) is far from a stable reference.
These figures directly undermine the assumption that dated, pinned model names resolve to a consistent serving configuration over time.

\Cref{tab:system-fingerprint-overlap} reports pairwise overlap statistics for all model pairs with non-empty fingerprint intersection.
We compute the Jaccard similarity over the set of unique fingerprint strings observed per model name, with the coverage figures reporting the proportion of each model name's unique-fingerprint set that lies in the intersection.

\begin{table*}[h]
\centering
\begin{tabular}{llS[table-format=3,table-number-alignment=right]S[table-format=1.2,table-number-alignment=right]S[table-format=2.1\%,table-number-alignment=right]S[table-format=2.1\%,table-number-alignment=right]}
\hline
Alias & Pinned model & {Shared \sysfpSh} & {Jaccard} & {A covered} & {B covered} \\
\hline
\texttt{gpt-4.1} & \texttt{gpt-4.1-2025-04-14} & 156 & 0.68 & 70.0\% & 95.7\% \\
\texttt{gpt-4o} & \texttt{gpt-4o-2024-08-06} & 121 & 0.83 & 88.3\% & 94.0\% \\
\hline
\end{tabular}
\caption{Non-empty cross-model \sysfp overlaps. Jaccard similarity is computed over the unique \sysfp sets associated with each model name.}
\label{tab:system-fingerprint-overlap}
\end{table*}

The most pronounced overlap occurs between \texttt{gpt-4o} and \linebreak \texttt{gpt-4o-2024-08-06}, which share 121 unique fingerprints, corresponding to 88.3\% of the unpinned set and 94.0\% of the pinned snapshot.
This is consistent with alias routing to the August 2024 snapshot during the collection window, as documented by OpenAI.
The \texttt{gpt-4.1} family exhibits an analogous structure: the generic alias shares 156 fingerprints with \texttt{gpt-4.1-2025-04-14}, covering 70.0\% and 95.7\% of the respective sets.
By contrast, \texttt{gpt-4o-2024-05-13} and \texttt{gpt-4o-2024-11-20} share no fingerprints with any other mo\-del name in the survey.

\begin{figure}[h]
\centering
\renewcommand{\arraystretch}{2.2}
\setlength{\tabcolsep}{3pt}

\begin{tabularx}{0.46\textwidth}{@{}p{1.1cm}
>{\centering\arraybackslash}X
>{\centering\arraybackslash}X
>{\centering\arraybackslash}X
>{\centering\arraybackslash}X
>{\centering\arraybackslash}X
>{\centering\arraybackslash}X
@{}}
    & \footnotesize gpt-4.1
    & \footnotesize gpt-4.1-2025-04-14
    & \footnotesize gpt-4o
    & \footnotesize gpt-4o-2024-05-13
    & \footnotesize gpt-4o-2024-08-06
    & \footnotesize gpt-4o-2024-11-20
    \\
\footnotesize gpt-4.1 \phantom{2025-04-14} & \cellcolor{T-D-PG8}1.00 & \cellcolor{T-D-PG6}\makecell{0.68\\n=156} & & & & \\
\footnotesize gpt-4.1-2025-04-14 & \cellcolor{T-D-PG6}\makecell{0.68\\n=156} & \cellcolor{T-D-PG8}1.00 & & & & \\
\footnotesize gpt-4o \phantom{-2025-04-14} & & & \cellcolor{T-D-PG8}1.00 & & \cellcolor{T-D-PG7}\makecell{0.83\\n=121} & \\
\footnotesize gpt-4o-2024-05-13 & & & & \cellcolor{T-D-PG8}1.00 & & \\
\footnotesize gpt-4o-2024-08-06 & & &\cellcolor{T-D-PG7}\makecell{0.83\\n=121} & & \cellcolor{T-D-PG8}1.00 & \\
\footnotesize gpt-4o-2024-11-20 & & & & & & \cellcolor{T-D-PG8}1.00 \\
\end{tabularx}
\caption{Pairwise Jaccard similarity of unique \sysfp values across requested model names, with shared fingerprint cardinality noted as $n=$.}
\Description{Table noting that \sysfp for \texttt{gpt-4.1} and \texttt{gpt-4.1-
2025-04-14} have a Jaccard similarity of 0.68 using 156 fingerprints; \textttt{gpt-4o} and \textttt{gpt-4o-2024-08-06} have a Jaccard similarity of 0.83 using 121 fingerprints}
\label{fig:system-fingerprint-overlap}
\end{figure}

These results establish that provider-supplied fingerprints both rotate and drift continuously under fixed model names (including pinned, dated snapshots), ruling \sysfp out as a stable reference for evaluation claims.
\section{Architectures for Referential Security}
\label{sec:architectures}

The evidence in \Cref{sec:limitations} shows that provider-supplied metadata does not currently provide referential stability.
We now discuss two architectures that could recover referential security:
\begin{enumerate*}
\item cryptographic attestation and 
\item external behavioral fingerprinting.
\end{enumerate*}
Each is a candidate resolution $r$ as per \Cref{def:resolution}, engineering stability over a context $C$ that includes some subset of the threats from \Cref{sec:threat-model}.
The two architectures differ in their trust assumptions, their deployment costs, and their robustness to adversarial updates.

\subsection{Cryptographic attestation}
\label{sec:attestation}

The first architecture shifts the problem inside the provider's infrastructure.
A provider runs model inference within a Trusted Execution Environment (TEE) and commits to the model weights and serving configuration at initialization
time.
Outputs are signed with a key that is hardware-attested to the enclave, and a
verifiable attestation report binds the signing key to the specific software
and hardware configuration.
A relying party (an auditor, a regulator, or a research replicator) can
verify that a signed response was produced by a specific configuration without
access to the model weights themselves.
The attestation report functions as the AI analogue of a content hash from \Cref{subsec:references-in-computing}.
Any change to the served configuration produces a different report, so the resolution from identifier to configuration satisfies $\mathit{stable}(r, i, C)$ by construction in the sense of \Cref{def:stability}.

More expressive variants replace signatures with zero-knowledge proofs. 
A provider can prove properties of a computation, for example that a given model hash was used or that inference operated within declared parameter bounds, without revealing the model itself~\cite{chen2024zkml,sun2024zkllm}.
These approaches are the subject of active research in the cryptographic ML community~\cite{peng2025survey}. 

This architecture assumes a provider who is trusted to set up the TEE correctly and to operate it honestly, but who cannot tamper with attested outputs post-setup without detection.
It does not protect against a provider who controls the attestation infrastructure end-to-end.
Under this trust assumption, attestation defends against all three threats from \Cref{sec:threat-model}: routine evolution, context-dependent serving, and adversarial concealed update.
A configuration change of any kind produces a different attestation report, and the provider cannot engineer indistinguishability at the output layer because the binding is at the configuration layer.
The principal limitation is deployment cost.
TEE-based inference at commercial scale requires hardware support, significant engineering effort, and a key management infrastructure that does not currently exist for AI systems.
Critically, the architecture requires active participation by the provider and is therefore unavailable to parties who need referential security \textit{without} provider cooperation.
Claims made about an attested configuration are therefore bound in the sense of \Cref{def:bound-claim} for any context $C$ in which the attestation infrastructure remains intact.

\subsection{External behavioral fingerprinting}
\label{sec:external}

The second architecture requires no provider cooperation.
An external party characterizes a model by probing its API and collecting a fingerprint: a compact representation of the model's input-output behavior that is stable under the model's identity and discriminative across distinct models.
Later parties can compare a new interaction against the stored fingerprint to assess whether the same model is being queried.

Fingerprinting weakens \Cref{def:stability} from a deterministic relation to a probabilistic one. The resolution holds with some confidence that grows with probe budget rather than holding by construction. This is a different mode of stability from that in \Cref{subsec:references-in-computing}. 
In comparison, hashes are stable by construction while DOIs and Git tags are stable because a maintainer commits to keeping them stable. 
Fingerprinting supplies stability as an inference from accumulated behavioral evidence, not as a property of the identifier or a commitment by the resolver.

This is the setting in which the \sysfp results in~\Cref{sec:limitations} are most directly informative.
The serving infrastructure variance visible there represents noise that any external fingerprinting scheme must tolerate without treating as evidence of model change.
Existing black-box fingerprinting methods were not designed for this tolerance requirement, and the level of configuration variance we observe is sufficient to defeat approaches that treat output distributions as direct model proxies.

The threat model for external fingerprinting is strictly weaker than for cryptographic attestation.
The provider is untrusted and the fingerprint is a probabilistic rather than cryptographic claim.
Against the non-adversarial threats from \Cref{sec:threat-nonadv}, fingerprinting offers \textit{detection power}: routine evolution and
context-dependent serving produce behavioral changes that a well-designed
fingerprint will eventually surface, even if any single probe is inconclusive.

Against adversarial concealed update~(\Cref{sec:threat-adv}), naive fingerprinting cannot provide proof of identity.
A provider who can identify the verifier's probes can tune the new configuration to remain indistinguishable on those probes, and no probe-based method can defeat an adversary who has optimized against it. 
Fingerprinting therefore offers detection of unintentional change while \textit{proof of identity} against a motivated adversary is considerably more difficult. 

To prevent provider identification, fingerprinting extraction will need to blend in with normal user traffic, diversify geolocation, IP addresses, tokens, user accounts, usage patterns and other access patterns a provider could use. 
The probe set itself faces a further constraint. 
Public probes allow third parties to replicate a fingerprint check, which is essential for reproducibility and audit, but also expose the probes to the provider. 
A workable scheme therefore splits the probe set in two:
\begin{enumerate*}
\item a public subset supports open verification;
\item a private subset, rotated over time, serves as a canary.
\end{enumerate*}
A provider that has adapted to the public probes will produce stable fingerprints on the public, while the private probes register a change. This divergence is then used to flag the adaptation.

\subsection{Choosing between architectures}
\label{sec:choice}

The two architectures are not substitutes.
Cryptographic attestation provides strong, verifiable identity claims and covers the full threat model, including adversarial concealed update, but requires provider participation and significant infrastructure investment.
It is most appropriate for regulatory compliance contexts in which providers have an incentive to cooperate with auditors.
External fingerprinting is deployable today without provider cooperation. 
It provides weaker probabilistic guarantees and covers only the non-adversarial threats, but it does so on infrastructure that already exists.
It is most appropriate for research reproducibility contexts in which the goal
is to flag likely configuration drift rather than to prove identity against a
sophisticated adversary.

A complete infrastructure for referential security likely requires both
\begin{enumerate*}
\item attestation for high-stakes regulatory decisions where cryptographic proof is
necessary, and
\item external fingerprinting as a continuous monitoring layer that
can detect drift without waiting for a formal audit cycle.
\end{enumerate*}
The workflows in \Cref{sec:workflows} map naturally onto this division,
with the relevant difference being the context $C$ over which stability is
required.
Audit validity and cross-provider equivalence require stability over a context that includes a provider with an incentive to conceal change, which only attestation can guarantee.
Reproducible science requires stability in a non-adversarial context (\Cref{sec:threat-nonadv}) and is well-served by fingerprinting as a
lightweight signal that a replication attempt is operating on a changed system.

\section{Conclusions}\label{sec:conclusions}

Current AI safety evaluation operates under an incomplete paradigm.
It measures substantive model behavior while leaving a prior referential condition unexamined.
In conventional software, security claims are bound to precise configurations through versions and cryptographic hashes.
The AI security and safety community has inherited this assumption without inheriting the artifacts that supported it.
Public model names play the role versions and hashes played in conventional software, but they do not behave the same way.
A name persists while weights, safety filters, and serving infrastructure shift beneath it.
The result is that audits, evaluations, and emerging regulations measure ephemeral states while assuming a referential stability that hosted AI systems do not provide.
To address this gap, we propose referential security as a new paradigm for AI evaluation.
Under this paradigm, model identity is treated as an empirically verifiable property rather than a nominal, provider-issued assertion, distinct from any substantive security claims it conditions.
This shift allows regulation to reach the systems behind product names, not just the names themselves.

\bibliographystyle{ACM-Reference-Format}
\bibliography{references.bib}

\end{document}